\documentclass{ws-procs975x65}

\usepackage{epsfig}

\chardef\quoteleftcode=\catcode96	% remember catcode of `
\newread\epsffilein    % file to \read
\newif\ifepsffileok    % continue looking for the bounding box?
\newif\ifepsfbbfound   % success?
\newif\ifepsfverbose   % report what you're making?
\newdimen\epsfxsize    % horizontal size after scaling
\newdimen\epsfysize    % vertical size after scaling
\newdimen\epsftsize    % horizontal size before scaling
\newdimen\epsfrsize    % vertical size before scaling
\newdimen\epsftmp      % register for arithmetic manipulation
\newdimen\pspoints     % conversion factor
\def\epsfbox#1{\global\def\epsfllx{72}\global\def\epsflly{72}%
   \global\def\epsfurx{540}\global\def\epsfury{720}%
   \def\lbracket{[}\def\testit{#1}\ifx\testit\lbracket
   \let\next=\epsfgetlitbb\else\let\next=\epsfnormal\fi\next{#1}}%
\def\epsfgetlitbb#1#2 #3 #4 #5]#6{\epsfgrab #2 #3 #4 #5 .\\%
   \epsfsetgraph{#6}}%
\def\epsfnormal#1{\epsfgetbb{#1}\epsfsetgraph{#1}}%
\def\epsfgetbb#1{%
\openin\epsffilein=#1
\ifeof\epsffilein\errmessage{I couldn't open #1, will ignore it}\else
   {\epsffileoktrue \chardef\other=12
    \def\do##1{\catcode`##1=\other}\dospecials \catcode`\ =10
    \catcode`\^^L=9 \catcode`\^^?=9
    \loop
       \read\epsffilein to \epsffileline
       \ifeof\epsffilein\epsffileokfalse\else
	  \expandafter\epsfaux\epsffileline:. \\%
       \fi
   \ifepsffileok\repeat
   \ifepsfbbfound\else
    \ifepsfverbose\message{No bounding box comment in #1; using defaults}\fi\fi
   }\closein\epsffilein\fi}%
\def\epsfclipstring{}% do we clip or not?  If so,
\def\epsfsetgraph#1{%
   \epsfrsize=\epsfury\pspoints
   \advance\epsfrsize by-\epsflly\pspoints
   \epsftsize=\epsfurx\pspoints
   \advance\epsftsize by-\epsfllx\pspoints
   \epsfxsize\epsfsize\epsftsize\epsfrsize
   \ifnum\epsfxsize=0 \ifnum\epsfysize=0
      \epsfxsize=\epsftsize \epsfysize=\epsfrsize
      \epsfrsize=0pt
     \else\epsftmp=\epsftsize \divide\epsftmp\epsfrsize
       \epsfxsize=\epsfysize \multiply\epsfxsize\epsftmp
       \multiply\epsftmp\epsfrsize \advance\epsftsize-\epsftmp
       \epsftmp=\epsfysize
       \loop \advance\epsftsize\epsftsize \divide\epsftmp 2
       \ifnum\epsftmp>0
          \ifnum\epsftsize<\epsfrsize\else
	     \advance\epsftsize-\epsfrsize \advance\epsfxsize\epsftmp \fi
       \repeat
       \epsfrsize=0pt
     \fi
   \else \ifnum\epsfysize=0
     \epsftmp=\epsfrsize \divide\epsftmp\epsftsize
     \epsfysize=\epsfxsize \multiply\epsfysize\epsftmp
     \multiply\epsftmp\epsftsize \advance\epsfrsize-\epsftmp
     \epsftmp=\epsfxsize
     \loop \advance\epsfrsize\epsfrsize \divide\epsftmp 2
     \ifnum\epsftmp>0
	\ifnum\epsfrsize<\epsftsize\else
	   \advance\epsfrsize-\epsftsize \advance\epsfysize\epsftmp \fi
     \repeat
     \epsfrsize=0pt
    \else
     \epsfrsize=\epsfysize
    \fi
   \fi
   \ifepsfverbose\message{#1: width=\the\epsfxsize, height=\the\epsfysize}\fi
   \epsftmp=10\epsfxsize \divide\epsftmp\pspoints
   \vbox to\epsfysize{\vfil\hbox to\epsfxsize{%
      \ifnum\epsfrsize=0\relax
        \includegraphics{#1}%
      \else
        \epsfrsize=10\epsfysize \divide\epsfrsize\pspoints
        \includegraphics{#1}%
      \fi
      \hfil}}%
\global\epsfxsize=0pt\global\epsfysize=0pt}%
{\catcode`\%=12 \global\let\epsfpercent=%\global\def\epsfbblit{%BoundingBox}}%
\long\def\epsfaux#1#2:#3\\{\ifx#1\epsfpercent
   \def\testit{#2}\ifx\testit\epsfbblit
      \epsfgrab #3 . . . \\%
      \epsffileokfalse
      \global\epsfbbfoundtrue
   \fi\else\ifx#1\par\else\fi\fi}%
\def\epsfempty{}%
\def\epsfgrab #1 #2 #3 #4 #5\\{%
\global\def\epsfllx{#1}\ifx\epsfllx\epsfempty
      \epsfgrab #2 #3 #4 #5 .\\\else
   \global\def\epsflly{#2}%
   \global\def\epsfurx{#3}\global\def\epsfury{#4}\fi}%
\def\epsfsize#1#2{\epsfxsize}
\catcode`\`=\quoteleftcode		% restore previous catcode

\begin{document}

\title{SPHERICAL VOIDS IN NEWTON-FRIEDMANN UNIVERSE}

\author{R. TRIAY}

\address{Centre de Physique Th\'eorique
\footnote{Unit\'e Mixte de Recherche (UMR 6207) du CNRS, et des universit\'es Aix-Marseille I,
Aix-Marseille II et du Sud Toulon-Var. Laboratoire affili\'e \`a la FRUMAM (FR 2291).}\\
CNRS Luminy Case 907, 13288 Marseille Cedex 9, France\\
$^*$E-mail: triay@cpt.univ-mrs.fr\\\
www.cpt.univ-mrs.fr}

\author{H. H. FLICHE}

\address{LMMT\footnote{UPRES~EA 2596}, Fac. des Sciences et Techniques de St J\'er\^ome\\
av. Normandie-Niemen, 13397 Marseille Cedex 20, France\\
E-mail: henri-hugues.fliche@wanadoo.fr}

\begin{abstract}
The understanding of voids formation, which is at the origin of the foam like patterns in the distribution of galaxies within scale up to 100 Mpc, has become an important challenge for the large scale formation theory\cite{Peebles01}. Such a structure has been observed since three decades and confirmed by recent surveys\cite{SoneiraPeebles78, JoeveerEinasto78, KirshnerEtal81,Einasto02, RojasEtal04, CrotonEtal04, JaanisteEtal04, ConroyEtal05}.  Investigations has been performed  -- on their statistical properties by improving identification techniques\cite{SantiagoEtal06}, by exploring their formation process in a $\Lambda$ CDM model through N-body simulations\cite{GottloberEtal03, ShandarinEtal04, GoldbergVogeley04, Mathis04,Colberg04,WeygaertEtal04,PadillaEtal05}, by probing their origins\cite{OrdEtal05}; -- on the kinematics of giant voids\cite{KopylovKopylova02} and the dynamics by testing models of void formation\cite{BenhamidoucheEtal99,FriedmannPiran00,Bolejko06}. Herein, we investigate the effect of the cosmological constant $\Lambda$ on the evolution of a spherical void through an exact solution of Euler-(modified)~Poisson equations system (EPES)\cite{FlicheTriay06}. Let us remind that Friedmann-Lema\^{\i}tre models, which provide us with a suitable description of the universe at large scales (thanks to their stability with respect to linear perturbations\cite{LifchitzKhalatnikov63}), can be described within a Newtonian approach by means of EPES solutions for whom kinematics satisfy Hubble (cosmological) law. The  void consists of three distinct media~:  a {\it material shell\/}  (S) with null thickness and negligible tension-stress, an empty inside and a uniform dust distribution outside which expands according to Friedmann equation. We use a covariant formulation of EPES\cite{Souriau70, DuvalKunzle78}  for deriving the evolution with time of ${\rm S}$ acting as boundaries condition for the inside and outside media.
\begin{figure}
\begin{center}
\epsfig{file=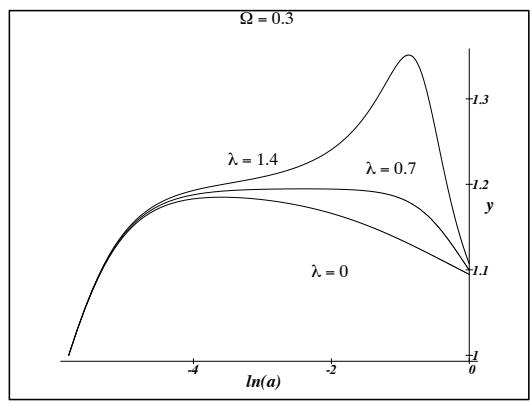,width=3in}
\end{center}
\caption{The corrective factor $y$ to Hubble expansion.
It results from a void that initially expands with Hubble flow at expansion parameter
$a_{\rm i}=0.003$, with $\Omega_{\circ}=0.3.$ and  $\Omega_{\lambda}=\lambda=0, 0.7, 1.4$.}
\label{Fig0}
\end{figure}

As a result, S expands with a huge initial burst that freezes asymptotically up to matching Hubble flow. The related perturbation on redshift of sources located on S does not exceed $\Delta\,z\sim 10^{-3}$. In the Friedmann comoving frame, its magnification increases nonlinearly with $\Omega_{\circ}$ and $\Lambda$. These effects interpret respectively by the gravitational attraction from the outer parts and repulsion (of vacuum) from the inner parts of S, with a sensitiveness on $\Omega$ at primordial epochs and on $\Lambda$ later on by preserving the expansion rate from an earlier decreasing. This dependence of the expansion velocity $\vec{v}=y H \vec{r}$ on $\Lambda$ is shown on Fig.\,\ref{Fig0} through the corrective factor  $y$ to Hubble expansion, where $\vec{r}$ and $H$ stand respectively for radius of S and Hubble parameter at time $t$. It is characterised by a protuberance at redshift $z\sim 1.7$, the larger the $\Lambda$ the higher the bump. It is due to the existence of a minimum value of Hubble parameter $H$  which is reached during the cosmological expansion (also referenced as a loitering period). It characterises spatially closed Friedmann models that expands for ever, they offer the property of sweeping out the void region, what interprets as a stability criterion.
\end{abstract}

\keywords{Cosmology : Theory, Cosmological Constant, Voids, Large Scale Structures.}

\end{document}